# Second-order nonlinear optical and linear UV-VIS absorption properties of type-II multiferroic candidates RbFe($A$O$_4$)$_2$ ($A$ = Mo, Se, S)

Rachel Owen[1], Elizabeth Drueke[1], Charlotte Albunio[1], Austin Kaczmarek[1], Wencan Jin[2], Dimuthu Obeysekera[3], Sang-Wook Cheong[4], Junjie Yang[3], Steven Cundiff[1], and Liuyan Zhao[1,*]

[1] *Physics Department, University of Michigan, 450 Church Street, Ann Arbor, MI, 48109 USA*
[2] *Department of Physics, Auburn University, 380 Duncan Drive, Auburn, AL, 36849, USA*
[3] *Department of Physics, New Jersey Institute of Technology, 323 Dr Martin Luther King Jr Blvd, Newark, NJ, 07102, USA*
[4] *Rutgers Center for Emergent Materials and Department of Physics and Astronomy, Rutgers University, Piscataway, NJ, 08854, USA*

[*] corresponding to lyzhao@umich.edu

Motivated by the search for type-II multiferroics, we present a comprehensive optical study of a complex oxide family of type-II multiferroic candidates: RbFe(MoO$_4$)$_2$, RbFe(SeO$_4$)$_2$, and RbFe(SO$_4$)$_2$. We employ rotational-anisotropy second harmonic generation spectroscopy (RA SHG), a technique sensitive to point symmetries, to address discrepancies in literature-assigned point/space groups and to identify the correct crystal structures. At room temperature we find that our RA SHG patterns rotate away from the crystal axes in RbFe($A$O$_4$)$_2$ ($A$ = Se, S), which identifies the lack of mirror symmetry and in-plane two-fold rotational symmetry. Also, the SHG efficiency of RbFe(SeO$_4$)$_2$ is two orders of magnitude stronger than RbFe($A$O$_4$)$_2$ ($A$ = Mo, S), which suggests broken inversion symmetry. Additionally, we present temperature-dependent linear optical characterizations near the band edge of this family of materials using ultraviolet-visible (UV-VIS) absorption spectroscopy. Included is experimental evidence of the band gap energy and band gap transition type for this family. Previously unreported sub-band gap absorption is also presented, which reveals prominent optical transitions, some with an unusual central energy temperature dependence. Furthermore, we find that by substituting the $A$-site in RbFe($A$O$_4$)$_2$ ($A$ = Mo, Se, S), the aforementioned transitions are spectrally tunable. Finally, we discuss the potential origin and impact of these tunable transitions.

## I. INTRODUCTION

Ferroics comprise a large class of materials that rarely share coupled electric and magnetic order parameters [1-4]. This can be understood through the Landau theory of phase transitions, which demonstrates that magnetism and ferroelectricity break time-reversal (TR) and spatial-inversion (SI) symmetry, respectively [5,6]. Since magnetism and ferroelectricity break different symmetries, rarely do they couple with one another linearly. Multiferroics are a class of atypical ferroic materials where magnetism and ferroelectricity coexist. Realized in a multitude of materials, type-I multiferroics typically demonstrate a linear magnetoelectric response and tend to have strong ferroelectric polarizations [7]. A common example of a type-I multiferroic is bismuth ferrite (BiFeO$_3$), which is both ferroelectric ($T_{Curie}$ ~ 1103 K) and antiferromagnetic ($T_{Neel}$ ~ 643

K). In its thin film form, BiFeO$_3$ shows exceptionally strong spontaneous polarization and the magnetic and ferroelectric order parameters can couple even though the ferroelectric and magnetic transitions do not emerge jointly [8]. This material and other type-I multiferroics have drawn much attention for their potential applications in spintronics, sensors, and information storage. However, because the magnetic and ferroelectric transitions do not emerge jointly, type-I multiferroics generally have weak magnetoelectric coupling. This has stimulated a search for type-II multiferroics, which are defined by strong magnetoelectric coupling derivative from processes in which the magnetic order induces the electric order.

One such discovered type-II multiferroic is the complex oxide, RbFe(MoO$_4$)$_2$, which not only has strong magnetoelectric coupling effects but is also a rare example of a quasi 2D-antiferromagnet on a triangular planar lattice (2D-TLA) below 3.8 K [9,10]. As such, RbFe(MoO$_4$)$_2$ has attracted much attention from both the multiferroics and quantum magnetism communities. While studies have predominantly focused on magnetic properties in RbFe(MoO$_4$)$_2$, there has also been significant work to determine the room temperature space group and observe the predicted ferro-rotational ordered phase transition from $P\bar{3}m1$ to $P\bar{3}$ at critical temperature $T_C$ = 195 K [11-17]. A recent second harmonic generation (SHG) study has successfully identified this ferro-rotational order and shown its physical properties such as uneven domain distribution and nontrivial coupling fields [11].

While second-order nonlinear optical processes in RbFe(MoO$_4$)$_2$ have been studied, to our knowledge, basic linear optical properties have yet to be determined for this material. These optical properties provide useful information about a material such as the band gap energy, optical transition type (direct or indirect), and presence of electronic states. Density functional theory (DFT) calculations have revealed RbFe(MoO$_4$)$_2$ to be a wide-band gap semiconductor with relatively flat valence and conduction bands [18], making it difficult to distinguish the band gap as being direct or indirect without experimental investigations [19]. Interesting linear optical properties of some multiferroic complex oxides have also demonstrated relevance to applications such as tunable solar cells [20]. Aside from supporting the understanding of the nonlinear optical processes in this material, the examination of the linear optical properties could independently motivate future studies and applications.

The aims of our study are two-fold. First, we focus on determining the basic optical properties of RbFe(MoO$_4$)$_2$ to characterize the valence-conduction band transition and to determine the presence of any additional electronic states. Second, we use this information to aid in the widespread search for multiferroics with interesting ferro-rotational orders. The rotation between the FeO$_6$ octahedra and MoO$_4$ tetrahedra, which can be seen when comparing the RbFe(MoO$_4$)$_2$ and RbFe(SO$_4$)$_2$ crystal diagrams in Fig. 1 (a), is responsible at lower temperatures for the ferro-rotational ordering in RbFe(MoO$_4$)$_2$. This rotation or twisting is a prerequisite for the multiferroic ordering at very low temperatures. Thus, we can gain new insight to both the ferro-rotational and multiferroic properties by replacing the molybdenum site and exploring how the symmetry and band structure are affected.

Two promising candidates for interesting ferro-rotational ordering that obey the stacking structure of RbFe(MoO$_4$)$_2$ are immediately apparent. One is RbFe(SO$_4$)$_2$, which has been predicted by DFT and shown by X-ray diffraction (XRD) measurements to be of the point group 32 at room temperature [18,21]. Conversely, neutron diffraction measurements have shown RbFe(SO$_4$)$_2$ to be either $\bar{3}$ or $\bar{3}m$ [22,23]. The other candidate is the largely unstudied RbFe(SeO$_4$)$_2$, which is predicted by DFT and shown by XRD to be in the point group 32 at room temperature [18,24]. Important to note is that the relationship between the $AO_4$ ($A$ = S, Se) tetrahedra and the FeO$_6$ octahedra is similar among both candidate materials and RbFe(MoO$_4$)$_2$. This indicates that studying these materials could result in insight about their ferro-rotational and multiferroic properties as well as determine the presence of any interesting optical transitions.

In this study, we aim to give additional insight to variation in the crystal structure and second-order nonlinear optical transitions among the complex oxide family RbFe(*A*O$_4$)$_2$ *A* = (Mo, Se, S) using rotational-anisotropy (RA) SHG spectroscopy. Additionally, we compare the linear optical properties of these three materials at room temperature and investigate their temperature dependence. We present experimental estimations for the band gap energies in these wide band gap semiconductors, show insight into the type of optical transition between the valence and conduction bands, and demonstrate previously unreported sub-band gap optical transitions caused by in-gap electronic states. In section II, we describe the various growth methods for these single crystals along with sample preparations for optical measurements. In section III, we employ RA SHG spectroscopy to determine the precise point group of these materials to overcome the inherent systematic absences in crystallographic methods. We compare experimental measurements to simulated RA SHG patterns based on point groups suggested in literature. In section IV, we show temperature dependent UV-VIS absorption measurements. We discuss the temperature dependence of the band edge in addition to presenting in-gap electronic states not yet reported in all three materials. Section V provides a summary of our findings.

## II. SAMPLE GROWTH AND PREPARATION FOR OPTICAL MEASUREMENTS

Growth methods of single crystals in this study vary between compounds. RbFe(MoO$_4$)$_2$ single crystals were synthesized using the flux melt method [11,12]. A powder mixture of Rb$_2$CO$_3$, Fe$_2$O$_3$, and MoO$_3$ (Alfa Aesar, 5 N purity) with the molar ratio 2:1:6 was heated in air in a platinum crucible at 1100 K for 20 h. The mixture was then cooled to 900 K at a rate of 2 K h$^{-1}$ followed by cooling to room temperature at a rate of 5 K h$^{-1}$. The resulting transparent light-green hexagonal platelet crystals with approximate dimensions of 3 × 3 × 0.1 mm$^3$ were then separated from the flux by dissolving in warm water.

The RbFe(SO$_4$)$_2$ and RbFe(SeO$_4$)$_2$ single crystals were both grown using a hydrothermal method [25]. A sulfuric (or selenic) acid aqueous solution of Rb$_2$SO$_4$ (or Rb$_2$SeO$_4$) and Fe$_2$(SO$_4$)$_3$ (or Fe$_2$(SeO$_4$)$_3$) with a molar ratio of 1:1 was sealed in a hydrothermal autoclave with a Teflon liner and kept in a furnace around 380 - 480 K for 72 h. Transparent clear hexagonal platelet crystals with approximate dimensions of 7 × 5 × 0.1 mm$^3$ for RbFe(SO$_4$)$_2$ and 3 × 3 × 1 mm$^3$ for RbFe(SeO$_4$)$_2$ were then separated from the solution.

For RA SHG measurements, as-grown single crystals were mounted to a stage in ambient conditions. Due to the layered nature of the materials, any uneven or loose top layers on the single crystals were removed prior to measurements using carbon tape. The cleavage of the samples is comparable to mica, especially for the case of RbFe(SeO$_4$)$_2$ which separates into individual layers preserving the entire area of the hexagonal face.

To overcome penetration depth restrictions on the as-grown crystals for UV-VIS absorption measurements, the RbFe(MoO$_4$)$_2$ and RbFe(SO$_4$)$_2$ platelet crystals were further cleaved using a mechanical stress etching procedure. The RbFe(SeO$_4$)$_2$ crystals were separated into thin individual hexagonal layers using carbon tape. The samples were bonded to a transparent sapphire substrate during these processes and during the absorption measurements. The final surfaces for all three materials were then wiped clean using acetone, isopropyl alcohol, and methanol. Exact thicknesses of the prepared samples were not determined. However, using RbFe(SeO$_4$)$_2$ ellipsometry measurements and cutoff wavelengths from absorbance measurements, final thicknesses are estimated to be on the order of 1.5 µm (see Appendix D).

## III. NONLINEAR OPTICAL SPECTROSCOPY TO DETERMINE CRYSTAL STRUCTURES

### A. Rotational-anisotropy second harmonic generation spectroscopy

SHG, or frequency doubling, is a process in which the frequency of incident light is doubled through second-order light-matter interactions within a material. Traditionally, the measurement of SHG is used to determine the second-order nonlinear response of non-centrosymmetric crystals where the leading electric dipole (ED) contribution to the SHG is present. For materials with SI symmetry, highly sensitive detection schemes are required to measure higher-order SHG contributions such as the electric quadrupole (EQ) transition [26,27]. For this family of complex oxides, RbFe(SeO$_4$)$_2$ [18,24] and possibly RbFe(SO$_4$)$_2$ [18,21], are predicted to exhibit the leading-order ED processes due to broken inversion symmetry. The polarization in this case can be expressed as

$$P_i^{eff}(2\omega) = \chi_{ijk}^{ED} E_j(\omega) E_k(\omega) \tag{1}$$

where $\chi_{ijk}^{ED}$ is the ED second-order optical susceptibility tensor and the electric fields $E_j(\omega)$ and $E_k(\omega)$ correspond to the incident light. The next highest order contribution of the EQ transition to the SHG response must be considered for RbFe(MoO$_4$)$_2$ [11] and possibly RbFe(SO$_4$)$_2$ [22,23]. This EQ SHG response follows

$$P_i^{eff}(2\omega) = \chi_{ijkl}^{EQ} E_j(\omega) \partial_k E_l(\omega) \tag{2}$$

where $\chi_{ijkl}^{EQ}$ is now the EQ second-order optical susceptibility tensor. The forms of both $\chi_{ijk}^{ED}$ and $\chi_{ijkl}^{EQ}$ are determined by the crystal symmetry, while the absolute strengths of the tensor elements are material-specific.

RA SHG spectroscopy measures the SHG signal intensity, $I_{S_{in}-S_{out}}^{2\omega}(\phi)$, where $S_{in/out}$ can be substituted by $P_{in/out}$ leading to four possible polarization channels. $P/S_{in/out}$ corresponds to the incident/reflected (in/out) light being parallel/normal ($P/S$) to the light scattering plane. The angle $\phi$ is the azimuthal angle between the light scattering plane and the in-plane crystal axis in a selected polarization channel. For the oblique incidence geometry, the light is incident to the surface at an offset angle relative to the out-of-plane $c||z$ axis and the electric field contains both out-of-plane and in-plane components. This experimental geometry maximally accesses the second-order nonlinear susceptibility tensor elements. At normal incidence, where the electric field is parallel to the sample surface, only tensor elements without a $z$-component are probed. This reduces the number of possible polarization channels to two, which are called the parallel (counterpart of $S_{in}/S_{out}$ in the oblique incidence) and cross channels ($S_{in}/P_{out}$). A diagram of the experimental configuration can be found in Ref. [26].

To identify crystal structures, RA SHG spectroscopy is often used in conjunction with other crystallography techniques. For example, RA SHG is extremely sensitive to point symmetries but lacks sensitivity to translational symmetries. Diffraction techniques, on the contrary, excel at detecting translational symmetry but face challenges in capturing intra-unit cell point symmetries. This is due to systematic absences or extinctions when the structure factor is zero due to centered unit cells or the presence of glide or screw symmetry elements [28]. These systematic absences in crystallography can make certain space groups indistinguishable from each other within a specific crystal system. One example is difficulty distinguishing between space groups $P\bar{3}m1$, $P321$, and $P\bar{3}$ [22], which can explain the discrepancies in the literature for RbFe(SO$_4$)$_2$. This means additional techniques are needed to account for these discrepancies. Since RA SHG spectroscopy is highly sensitive to slight variations between point symmetries, such as mirror reflections and rotational symmetries, it is a useful method to account for systematic absences in crystallography techniques.

### B. Trigonal point group simulations

We know that each of the three materials has a trigonal lattice structure from XRD [13,14,21,24], but there are variations in specific point groups that RA SHG can differentiate and XRD cannot. This includes the presence of mirror planes normal to the layers or rotational axes within the layers. Since the structures of RbFe($A$O$_4$)$_2$ ($A$ = Mo, Se, S) belong to the trigonal crystal class, each has three-fold rotational symmetry about the out-of-plane $c$-axis ($C_3$). This can be seen in Fig. 1 (a) which depicts each predicted crystal structure at room temperature as viewed along the $c$-axis. Diagrams in Fig. 1 (a) were created using the open source crystal toolkit from Ref. [18] and the software VESTA [29].

As discussed, RA SHG spectroscopy techniques have confirmed that RbFe(MoO$_4$)$_2$ belongs to the point group $\bar{3}m$ at room temperature [11], consistent with the assigned space group of $P\bar{3}m1$ [13,14]. At room temperature, RbFe(SeO$_4$)$_2$ is predicted to belong to point group 32, which is symmorphic to the space groups $P321$ [18,24]. RbFe(SO$_4$)$_2$ is predicted to belong to one of point groups 32, $\bar{3}m$, or $\bar{3}$ which correspond to the space groups $P321$, $P\bar{3}m1$, and $P\bar{3}$, respectively [18,21-23]. Since the most recent results predict space group $P\bar{3}$, we use this for our diagram in Fig. 1 (a).

We derive the leading-order susceptibility tensors based on the above point groups and calculate the functional form of the RA SHG intensity for every material using Equations (1) and (2) (see Appendix A). For RbFe(MoO$_4$)$_2$, of the point group $\bar{3}m$, we describe the RA EQ SHG functional form for the parallel and cross channels at normal incidence as:

$$I^{2\omega}_{Parallel}(\phi) = \left(\chi^{EQ}_{yyzy}\cos(3\phi)\right)^2 \quad (3)$$

$$I^{2\omega}_{Cross}(\phi) = \left(\chi^{EQ}_{yyzy}\sin(3\phi)\right)^2 \quad (4)$$

For RbFe(SeO$_4$)$_2$, the symmetries of the point group 32 yield for the RA ED SHG intensity:

$$I^{2\omega}_{Parallel}(\phi) = \left(\chi^{ED}_{yyx}\sin(3\phi)\right)^2 \quad (5)$$

$$I^{2\omega}_{Cross}(\phi) = \left(\chi^{ED}_{yyx}\cos(3\phi)\right)^2 \quad (6)$$

For RbFe(SO$_4$)$_2$, the symmetries of the point group $\bar{3}$, yield for the RA EQ SHG intensity:

$$I^{2\omega}_{Parallel} = \left(\chi^{EQ}_{yyzy}\cos(3\phi) + \chi^{EQ}_{xxzx}\sin(3\phi)\right)^2 \qquad (7)$$

$$I^{2\omega}_{Parallel}(\phi) = \left(\chi^{EQ}_{xxzx}\cos(3\phi) - \chi^{EQ}_{yyzy}\sin(3\phi)\right)^2 \qquad (8)$$

where we note that the elements of the EQ susceptibility tensors are unique compared to those in Equations (3) and (4).

At this point, it is prudent to recognize that normal incident RA SHG measurements are sufficient to distinguish between the possible point groups of each material. To distinguish between ED SHG and EQ SHG, we compare the magnitude of the SHG response for each material. In addition, we can distinguish between the various trigonal point groups by comparing the orientation of the RA SHG patterns. We note that for point groups $\bar{3}m$ and 32 the RA SHG patterns are locked to the in-plane crystal axes but differ by 90° from one another. The patterns for $\bar{3}$ in contrast, can rotate off the crystal axes. Physically, this rotation corresponds to the opposing rotation of the FeO$_6$ octahedra with the MoO$_4$ tetrahedra as demonstrated for RbFe(MoO$_4$)$_2$ at lower temperatures [11].

### C. Crystal structure determination

For our RA SHG measurements, the incident fundamental light source has a wavelength of 800 nm, pulse duration of 40 fs, and a 200 kHz repetition rate. For RbFe(MoO$_4$)$_2$ and RbFe(SO$_4$)$_2$, the beam diameter at the sample was 25-50 μm with laser fluences of 0.25-0.75 mJ cm$^{-2}$. For RbFe(SeO$_4$)$_2$, the beam diameter was 1 μm to perform SHG scanning measurements with laser fluences of 15 mJ cm$^{-2}$. Relative signal levels shown in Fig. 1 are corrected to account for the differences in fluences and in experimental geometry. To determine the magnitude of the SHG response, we compare the effective susceptibility strength without Fresnel corrections (Appendix D) such that a signal level of 1 corresponds to an effective susceptibility strength of $\chi^{eff}_{yyy} = 8 \times 10^{-4}$ pm·V$^{-1}$ ($\chi^{eff}_{yyy} = \chi^{ED}_{yyy}$ for ED SHG and $\chi^{eff}_{yyy} = \chi^{EQ}_{yyzy}q_z$ for EQ SHG), which in this case corresponds to a power of 1 nW cm$^{-2}$ for an incidence light source with a fluence of 1 mJ cm$^{-2}$. To orient our measurements relative to the crystalline axes, white light or SHG scanning imaging is used to determine the angle offset between the crystal axes and the lab frame horizontal. The imaging detector and horizontal polarization are aligned to the lab frame. For the parallel channel RA SHG measurements shown in Fig. 1, the incident fundamental beam is vertically polarized at $\phi = 0°$.

The functional form for the RA SHG patterns in Equations (3) through (8) can be compared directly to the experimental measurements. For each material, the parallel and crossed channels were measured at normal incidence to the (001) or *ab*-crystal plane. The top row of Fig. 1 (b) shows, from left to right, the parallel channel RA SHG patterns at room temperature for RbFe(MoO$_4$)$_2$, RbFe(SeO$_4$)$_2$, and RbFe(SO$_4$)$_2$ respectively. To confirm that RbFe(SeO$_4$)$_2$ has broken inversion symmetry, we compare its SHG signal strength with the other two compounds in this study and its susceptibility tensor with known nonlinear crystals. RbFe(SeO$_4$)$_2$ consistently has an SHG signal level that is at least two orders of magnitude larger than the EQ SHG in RbFe(MoO$_4$)$_2$ [11]. The estimated SH susceptibility tensor element is $\chi_{yyy} \sim 0.45$ pm V$^{-1}$ with proper correction done using ellipsometry (see Appendix D) [30]. This value is comparable to those of similar frequency-doubling crystals [31]. These observations taken together motivate our assignment of the ED term as the leading contribution to the SHG response in RbFe(SeO$_4$)$_2$. Similarly, since RbFe(SO$_4$)$_2$ has an SHG signal level that is the same order of magnitude as RbFe(MoO$_4$)$_2$, we agree there is a center of inversion as predicted by neutron diffraction measurements [22] and therefore reject 32 [18,21] as a possible point group assignment.

After narrowing the possible point groups based on the dominant source of SHG (ED or EQ), we determine whether the RA SHG pattern originates from the expected point groups. The in-plane crystal axes are determined by the crystalline edges for RbFe(SeO$_4$)$_2$ and RbFe(SO$_4$)$_2$, which are confirmed by X-ray Laue diffraction measurements to be the *a*- and *b*- axis [25]. For RbFe(MoO$_4$)$_2$, the *a*- and *b*-axis are determined by oblique incidence RA SHG measurements [11]. Shown in Fig. 1 (b), both RbFe(MoO$_4$)$_2$ and RbFe(SO$_4$)$_2$ have RA SHG patterns consistent with the literature point group assignments of $\bar{3}m$ [13,14] and $\bar{3}$ [22,23] respectively. This can be seen by the RA SHG pattern alignment relative to the *a*-axis. The large rotation off of the *a*-axis for RbFe(SO$_4$)$_2$ resolves the ambiguity between the point group assignments of $\bar{3}m$ [22] and $\bar{3}$ [22,23] in neutron diffraction measurements of previous studies. The RA SHG pattern for RbFe(SeO$_4$)$_2$ is also rotated away from the crystal axes. This, however, is inconsistent with the literature-assigned point group of 32 for which the RA SHG patterns should be locked to the symmetry axes.

To account for this rotation, the RbFe(SeO$_4$)$_2$ RA SHG pattern is fit to the calculated RA SHG functional form for point group 3. Of the various trigonal point groups, 3 is the only one with an ED SHG contribution which simultaneously allows for rotation of the data off the crystal axes. Using the symmetries of the point group 3, we derive the functional forms for the ED SHG intensity for parallel and cross channel RA SHG measurements as:

$$I^{2\omega}_{Parallel}(\phi) = \left(\chi^{ED}_{yyy}\cos(3\phi) + \chi^{ED}_{yyx}\sin(3\phi)\right)^2 \quad (9)$$

$$I^{2\omega}_{Cross} = \left(\chi^{ED}_{yyx}\cos(3\phi) - \chi^{ED}_{yyy}\sin(3\phi)\right)^2 \quad (10)$$

A summarized comparison between the literature provided point groups and our assignments based on RA SHG spectroscopy can be found in the table in Fig. 1 (c). We have performed spatially-scanned RA SHG measurements on RbFe(SeO$_4$)$_2$ and find the rotation of the RA SHG pattern is consistent in both direction and magnitude throughout the sample (see Appendix C). Therefore, we are confident that the lowering of the crystal symmetry for RbFe(SeO$_4$)$_2$ is a uniform, global effect, instead of the result of some inhomogeneous local origin such as strain. Further diffraction measurements are needed to clarify the origin of this symmetry reduction.

## IV. LINEAR ABSORPTION SPECTROSCOPY TO PROBE ELECTRONIC STATES

### A. Absorption spectroscopy

After identifying the differences in crystal structure for the complex oxide family RbFe(*A*O$_4$)$_2$ *A* = (Mo, Se, S), we determine the presence of in-gap electronic states using UV-VIS absorption spectroscopy. To the best of our knowledge, there is no experimental study presenting sub-band gap absorption and the band gap energy of these three materials. Due to the flat nature of the band structure of RbFe(MoO$_4$)$_2$ and RbFe(SeO$_4$)$_2$, there is also ambiguity as to whether these wide band gap semiconductors have direct or indirect transitions, as valence band maxima and conduction band minima are difficult to determine [18]. As such, we compare different absorption models to estimate the band gap energy and provide additional insight to the type of band gap transition of these materials.

We employ transmission-based UV-VIS absorption spectroscopy due to the low reflectivity levels of the materials (see Appendix D). For the experimental set-up, the light source used was an Ocean Optics DH-2000 deuterium/halogen lamp with a wavelength range of 190-2500 nm (0.5 - 6.5 eV) with a multimode fiber-coupled power output of 217 µW. The lamp spectrum was further filtered to transmit wavelengths between 350-600 nm (2.07 - 3.54 eV). Single crystal thicknesses were reduced (see section II) such that sample transmission was detectable down to 370 nm (3.35 eV). An Ocean Optics Flame-S UV-VIS spectrometer was used with a detection range of 200 - 800 nm (1.55 - 6.20 eV). The spot size of the light source at the sample site was measured to have a full-width half max (FWHM) of 150 µm.

### B. Linear optical properties and electronic states

#### 1. UV-VIS room temperature results

UV-VIS absorbance measurements at room temperature for all three materials are presented in Fig. 2. One prominent feature in all three materials is the optical transition due to the presence of electronic states at energies 250 - 300 meV below the band gap. Additionally, it is noticeable that replacing the *A*-site with heavier elements for the complex oxide family RbFe($AO_4$)$_2$ $A$ = (Mo, Se, S) simultaneously decreases the band gap energy along with the peak energy of the in-gap electronic state. This atomic weight dependency is comparable to the tunability of other families of materials, such as CsPb$A_3$ $A$ = (Cl, Br, I) lead halide perovskites, where increasing the atomic weight of halide site results in smaller band gap energies [32].

To better quantify the material dependent linear optical response, we use standard fitting procedures to track changes in the band gap and peak energies of the in-gap electronic states. The band edge is estimated using the Tauc model which gives the relation between the photon energy and the band gap as

$$(\alpha \hbar\omega)^{1/n} = A(\hbar\omega - E_g) \quad (11)$$

where $\hbar\omega$ is the photon energy, $\alpha$ is the absorption coefficient, $E_g$ is the band gap, and $A$ is a proportionality constant [33]. The integer $n$ is determined by the type of optical transition. Typically, $n = 1/2$ for a direct allowed transition and $n = 2$ for an indirect allowed transition. Assuming low reflectivity levels, we use the absorbance measurements for the energy dependent absorption coefficient, $\alpha(\hbar\omega)$.

Comparing the fits for both $n = 1/2$ and $n = 2$, we find that the direct allowed transition model is a better fit across all three materials (see Appendix B). The Tauc plot using the direct transition model is shown in Fig. 2 (b) for each material. The low energy side of the optical transitions below the band gap are fit to a Gaussian assuming an inhomogeneous distribution of states. This choice of functional form is motivated by the fact that the Gaussian provides a better fit than a Lorentzian (for a homogeneous distribution) for the lower energy side of the peaks [34]. This spectral inhomogeneity is also noticeable from the high asymmetries present as shown in Fig. 2 (c). Finally, in Fig. 2 (d) we report values based on these fittings for the band gap and peak energies for all three materials relative to the atomic weight of the *A* site. Error bars for the band gap energy estimate are based on protocols from Ref. [35].

The estimated band gap energy from the absorption measurements is larger than that predicted by DFT in all three materials by 0.4 – 1 eV [18]. Current DFT predictions for these materials also indicate a tunability trend opposite to ours in which RbFe(MoO$_4$)$_2$ has the largest band gap energy and RbFe(SO$_4$)$_2$ has the lowest. Our experimentally estimated band gap energies can be used to correct these discrepancies.

### 2. Temperature dependence of in-gap electronic states

Using the fitting procedures shown in Fig. 2 (c), the temperature dependence of the electronic state optical transition below the band gap is tracked down to 5 K. Representative absorption spectra at selected temperatures for the in-gap electronic states are shown in Fig. 3 (a). In addition, peak fitting values shown are shown in Fig. 3 (b) and (c). When tracking the central peak energy, we observe that the resonance blue shifts for $A$ = (S, Se) and red shifts for $A$ = Mo at lower temperatures. The blue shifts are consistent with typical behavior of semiconductor exciton, defect, and impurity states because photons with energies below the transition energy can access these states through phonon assistance at higher temperatures. However, the observed red shift and change in temperature dependence behavior for after the structural phase transition at 190 K for $A$ = Mo is less consistent with this model. Above the phase transition, the blue shift indicates a phonon-assisted transition. Below the phase transition, the peak energy is red shifted and continues to red shift as the temperature is reduced.

One possible explanation for this behavior is the emergence of shallow trap states after the phase transition that cause the higher energy states in the inhomogeneous distribution to carry more spectral weight. Shallow trap states can emerge from the presence of defects, impurities, and/or structural distortions [36]. Such trap states have been reported to cause spectral red shifts with decreasing temperatures in lead halide perovskites, which are known to have strong structural distortion [37]. Studying the absorbance measurements at 295 K and 5 K in the first panel of Fig. 3 (a), we in fact see that the line shape becomes more asymmetric as the spectral weight of the low energy side is reduced. For the FWHM (Fig. 3 (c)), we find that there is broadening at higher temperatures in all three materials, consistent for both our trap state and phonon-assistance models [38].

The states shown in Fig. 3 occur at energies well below the band gap energy (250 – 350 meV) in the Urbach tail, denoted by the lines at lower energies in Fig. 2 (b). In many bulk semiconductors, states caused by defects or impurities are known to reside in this region [38]. RA SHG scanning measurements indicate that if defects or impurities are present, they are on the atomic level and are randomly distributed in the single crystals. We found indication of the presence of sites with an alternate crystal structure in the RbFe(SeO$_4$)$_2$. These rare sites occur on the scale of $< 1\ \mu m$ and are visible with scanning RA SHG measurements (see Appendix C). These sites occur infrequently, which implies the absence of large defective regions of our samples. Absorption spectroscopy is relatively sensitive to point defects and is less localized than methods typically used to determine lattice distortions such as transmission electron microscopy.

While largely consistent with defect-induced shallow levels, we note that we cannot attribute the exact origin of these electronic states below the band gap as either defect or exciton states based solely on our data. The difference between the peak energy and the band edge is atypical for exciton states in bulk semiconductors as they tend to lie closer to the band edge due to small binding energies. Here, the prominence of the peak at room temperature would imply an exciton binding energy of at least 25 meV. While other bulk wide-gap semiconductors such as GaN have reported high binding energies (20 meV) near room temperature, the location of theses exciton states is in disagreement with our absorption spectrum [39]. If the observed sub-band gap transitions are due to an exciton state, further techniques such as photoluminescence could be used to estimate potentially large binding energies. However, large binding energies on the order of hundreds of meV are usually found in dimensionally confined materials rather than in bulk material [40,41]. Even further, in the next section, we show that at lower temperatures additional peaks emerge between this resonance and the band gap energy, which additionally discredits the assignment of these peaks as exciton states.

### 3. Temperature dependence of band edge

At lower temperatures, additional electronic states manifest near the band edge in RbFe(MoO$_4$)$_2$ and RbFe(SO$_4$)$_2$. The temperature dependent absorbance measurements of these states and the band gap for all three materials are shown in Fig. 4 (a). These additional peaks are denoted by arrows at the 5 K absorbance spectrum in Fig. 4 (a). They are most prominent at 5 K and noticeably begin to blend into the background near 200 K. Estimated band gap energies and peak energies are presented in Fig. 4 (b) and (c). Only linear regimes were considered for fitting the band gap energy at lower temperatures. The temperature dependence of the band gap energy is shown in Fig. 4 (b) and is fit using a thermodynamic model from Ref. [42] given by

$$E_g(T) = E_g(0) - S\langle\hbar\omega\rangle(\coth(\hbar\omega/2kT) - 1) \qquad (12)$$

where $E_g(0)$ is the band gap at zero temperature, $S$ is a coupling constant, and $\langle\hbar\omega\rangle$ is the average phonon energy. This model was chosen over the more typical empirical Varshni relationship because, in addition to being more consistent with Huang-Rhys vibration modeling, this model is used to more reliably capture low temperature behavior [42].

    First, our observations of the band edge temperature dependence indicate that the band gap energies in all three materials tend to blue shift overall with decreasing temperature as predicted. There are, however, some noticeable discrepancies between the measurements and the thermodynamic model in Fig. 4 (b). One reason for these discrepancies could be attributed to the relatively large error from Tauc modeling and/or newly emerged features, which makes the Tauc model even less reliable in predicting band gap energies (see Appendix B) [35,43]. However, RbFe(SO$_4$)$_2$, which has the most agreement with Equation (12), arguably undergoes the most dramatic alteration to the line shape with a band gap energy blue shift of about 175 meV between 295 K and 5 K. In contrast, RbFe(SeO$_4$)$_2$ has the least number of features and is in least agreement with Equation (12) due to the redshift in the band gap energy below 80 K. The band gap energy of RbFe(MoO$_4$)$_2$, while consistent with Equation (12) at low temperatures, undergoes a noticeable blue shift near the phase transition temperature similar to the temperature dependence of the peak in Fig. 3 (a). This information indicates that an alternative model in which more complex effects compete with phonon broadening may be required, such as exploring the possibility of trap states or structural distortion in the materials.

    Second, additional electronic states emerge near the band edge in both RbFe(MoO$_4$)$_2$ and RbFe(SO$_4$)$_2$ below room temperature. There are two prominent peaks whose energies are shown in Fig. 4 (c) and a single peak at 3.18 eV at 5 K in RbFe(SeO$_4$)$_2$. The RbFe(MoO$_4$)$_2$ and RbFe(SO$_4$)$_2$ peak energies reside below the estimated band gap energies, while the RbFe(SeO$_4$)$_2$ peak resides above the estimated band gap energy. Studying Fig. 4 (c), we find that there is little to no

temperature dependence of the RbFe(MoO$_4$)$_2$ peaks for 0 – 80 K. The RbFe(SO$_4$)$_2$ peaks have a blue shift consistent with the thermodynamic model discussed in Section IV.B.2. In conjunction with the earlier discussion of shallow trap states, the lack of a blue shift for RbFe(MoO$_4$)$_2$ might be explained by competing effects such as those between phonon and trap states. For RbFe(MoO$_4$)$_2$ and RbFe(SO$_4$)$_2$, the FWHM of these peaks is noticeably larger than for the peak residing far below the band gap. The temperature dependence of the FWHM is also consistent with the other peaks shown in Fig. 3.

For the prominent spectral peaks in RbFe(MoO$_4$)$_2$ and RbFe(SO$_4$)$_2$, the origin could be as trivial as additional defect states in the crystals. A more interesting possibility, however, is the presence of exciton states. This suggestion does not compete with theories on trap states or structural distortions, as exciton states can be impacted by them. DFT modeling in conjunction with our absorption measurements indicate that RbFe(MoO$_4$)$_2$ is a direct band gap material [18]. If this is true, we might expect to observe exciton states in the absorption spectrum near the band edge, and both the proximity of the peaks to the band edge and their emergence at low temperatures is more indicative of exciton absorption features. For RbFe(SO$_4$)$_2$, there are no current DFT predictions of the band structure to our knowledge, so our assignment for a direct band gap material is solely based on our absorption measurements. DFT predicts RbFe(SeO$_4$)$_2$ to be an indirect band gap material with flat bands similar to RbFe(MoO$_4$)$_2$, which contrasts with our Tauc model fittings (Appendix B) [18]. If true DFT is correct, this could explain the absence of any additional peaks below the band edge. Ultimately, since the behavior of these resonances based on our absorbance measurements are not unique to exciton states, confirmation of the origin of these in-gap electronic states will require further studies.

## V. CONCLUSION

We employ RA SHG spectroscopy to resolve discrepancies in literature-assigned point groups in the complex oxide family, RbFe($A$O$_4$)$_2$ $A$ = (Mo, Se, S). RbFe(MoO$_4$)$_2$ is a reported type-II multiferroic and 2D-TLA with multiple studies reporting consistent point group assignments. RbFe(SO$_4$)$_2$ is also a reported 2D-TLA, yet there are significantly fewer studies that include determination of point groups for either RbFe(SO$_4$)$_2$ or RbFe(SeO$_4$)$_2$. Since RA SHG spectroscopy is sensitive to point symmetries, we can account for systematic absences in crystallographic techniques such as XRD that make distinguishing between certain point groups challenging.

In agreement with previous work, we find RbFe(MoO$_4$)$_2$ to belong to the point group $\bar{3}m$ at room temperature [11]. We confirm that RbFe(SeO$_4$)$_2$ breaks inversion symmetry such that the ED transition is the leading contribution to the SHG. However, we find RbFe(SeO$_4$)$_2$ does not have three two-fold ($C_2$) rotational symmetry. As such, we assign RbFe(SeO$_4$)$_2$ to belong to the point group 3 at room temperature, challenging the DFT and crystallography assignment of 32 [18,24]. RbFe(SO$_4$)$_2$ has been assigned to point groups 32, $\bar{3}m$, and $\bar{3}$ by various studies [18,21-23]. We find our results align with the assignment of $\bar{3}$ from neutron diffraction results. We confirm that the EQ transition is the leading order contribution to the SHG, ruling out 32. Additionally, we demonstrate a lack of in-plane two-fold ($C_2$) rotational symmetry axes, ruling out $\bar{3}m$. This is of particular interest as it indicates RbFe(SO$_4$)$_2$ has the same point group at room temperature as RbFe(MoO$_4$)$_2$ below 195 K where ferro-rotational ordering is present [11]. Future studies may include temperature dependent RA SHG measurements to further analyze these off-axis rotations between the FeO$_6$ octahedra and $A$O$_4$ tetrahedra.

This study also presents experimental linear optical characterizations for the family RbFe($A$O$_4$)$_2$ $A$ = (Mo, Se, S) using UV-VIS transmission absorption spectroscopy. By employing the Tauc method [33,36], we report band gap energies for these wide band gap semiconductors at room temperature and low temperature and predict all three to have a direct band gap transition. We find that this family has a tunable band gap, where the atomic weight of the $A$-site is anti-corollary to the band gap energy. We discover the presence of multiple unreported sub-band gap optical transitions due to in-gap electronic states in all three materials and remark on potential origins based on temperature dependent behavior.

The first electronic state discussed occurs 250 – 350 meV below the band edge in all three materials and is assigned to likely be due to defect or impurity states. The central energy of these transitions has similar tunability as the band gap energy with regards to the atomic weight of the $A$-site. The temperature dependence of this electronic state in RbFe(MoO$_4$)$_2$ also uniquely shows interesting behavior consistent with a material possessing shallow trap states. The second set of electronic states discussed occur close to the band edge in RbFe(MoO$_4$)$_2$ and RbFe(SO$_4$)$_2$. These states, while possibly more defects states, have the potential to be exciton states and require further investigation as absorption spectroscopy is not sufficient to distinguish the origin of the spectral features.

Additional studies are proposed to determine the nature of these states using techniques such as photoluminescence and vibrational spectroscopy [36]. Regardless, the presence of any in-gap electronic states could have implications for ferroelectric properties as they affect the polarization of the material [44]. If defects or impurities are present, one possibility is less efficient coupling between any ferroelectric and ferromagnet orders at lower temperatures, which can be addressed through fabrication techniques [45].

## ACKNOWLEDGEMENTS

We knowledge helpful discussions with Mathew Day from the University of Michigan. Ellipsometry measurements presented in the appendix were performed in part at the University of Michigan Lurie Nanofabrication Facility. L. Zhao acknowledges support by National Science Foundation CAREER Award No. DMR-1749774. E.Drueke acknowledges support by the NSF Graduate Research Fellowship Program under grant No. DGE-1256260. The work at Rutgers University was supported by the DOE under Grant No. DOE: DE-FG02-07ER46382.

## APPENDIX

## APPENDIX A: TRIGONAL POINT GROUP SIMULATIONS

The RA SHG patterns at normal incidence are fit to the calculated functions given in section III. B. The forms of the intensity as a function of rotation angle $\phi$ for EQ SHG and ED SHG are given respectively by

$$I^{2\omega}(\phi) = \left|A\hat{e}_i(2\omega)\chi_{ijk}^{ED}\hat{e}_j(\omega)\hat{e}_k(\omega)\right|^2 I^\omega I^\omega$$

$$I^{2\omega}(\phi) = \left|A\hat{e}_i(2\omega)\chi_{ijkl}^{EQ}\hat{e}_j(\omega)\hat{\partial}_k(\omega)\hat{e}_l(\omega)\right|^2 I^\omega I^\omega$$

where $A$ is a constant determined by experimental geometry, $\hat{e}_i$ is the polarization of the incoming fundamental beam or outgoing SHG, $\chi^{ED}_{ijk}$ and $\chi^{EQ}_{ijkl}$ are the bulk ED and EQ susceptibility tensors, respectively, $\hat{\partial}_k \rightarrow \hat{q}_k$ where $\hat{q}_k$ is the wavevector of the incident fundamental light, and $I^\omega$ is the intensity of the incident beam. Experimentally, the polarization of the incident fundamental beam is rotated such that the rotation angle $\phi$ corresponds to rotating the sample perpendicular to the scattering plane at normal incidence.

### A. EQ SHG under $\bar{3}m$ point group

The functional form of the EQ SHG intensity under the $\bar{3}m$ point group is given by Equations (3) and (4) in section III. B. The point group $\bar{3}m$ in addition to a three-fold rotational symmetry axis about the out-of-plane $c$-axis ($C_3$) has a center of inversion, three two-fold ($C_2$) rotational symmetry axes along the in-plane $a$-axis and every 60° in-plane about the $c$-axis, $S_6$ rotations about the $c$-axis, and three $\sigma_d$ reflections with axes perpendicular to the $a$-axis and again every 60° in-plane about the $c$-axis. Using the crystal and experimental symmetries (interchangeable incident electric fields), we determine the indices ($ijkl$) of the eleven non-zero independent elements of $\chi^{EQ}_{ijkl}$: $yyyy = xxxx = yxyx + yxxy + yyxx$; $yyxx = xxyy = yxxy = xyyx$; $xxzz = yyzz = xzzx = yzzy$; $zzxx = zzyy = zxxz = zyyz$; $yyyz = -yxxz = -xyxz = -xxyz = yzyy = -yzxx = -xzyx = -xzxy$; $yyzy = -yxzx = -xyzx = -xxzy$; $zyyy = -zyxx = -zxyx = -zxxy$; $yxyx = xyxy$; $xzxz = yzyz$; $zxzx = zyzy$; $zzzz$.

### B. EQ SHG under $\bar{3}$ point group

The functional form of the EQ SHG intensity under the point group $\bar{3}$ is given by Equations (7) and (8) in section III. B. The point group $\bar{3}$, in addition to three-fold rotational symmetry, has a center of inversion, and an $S_6$ rotation about the $c$-axis. Using the crystal and experimental symmetries, we determine the indices for the eighteen non-zero independent elements of $\chi^{EQ}_{ijkl}$: $yyyy = xxxx = yyxx + yxxy + yxyx$; $yyxx = xxyy = xyyx = yxxy$; $xyxy = yxyx$; $yyzz = xxzz = yzzy = xzzx$; $zzyy = zzxx = zyyz = zxxz$; $yzyz = xzxz$; $zyzy = zxzx$; $yxzz = yzzx = -xyzz = -xzzy$; $zzyx = zxyz = -zzxy = -zyxz$; $xzyz = -yzxz$; $yyyx = yxyy = -xxxy = -xyxx = xxyx + xyxx + yxxx$; $xxyx = -yyxy$; $xyyy = -yxxx$; $yyyz = yzyy = -yxxz = -yzxx = -xyxz = -xzxy = -xxyz = -xzyx$; $xxxz = xzxx = -yzyx = -yxyz = -yzxy = -yyxz = -xyyz = -xzyy$; $xxzx = -xyzy = -yyzx = -yxzy$; $zxxx = -zyxy = -zxyy = -zyyx$; $yyzy = -yxzx = -xyzx = -xxzy$; $zyyy = -zxyx = -zyxx = -zxxy$; $zzzz$.

### C. ED SHG under 32 point group

The functional form of the ED SHG intensity under point group 32 is given by Equations (5) and (6) in section III. B. The point group 32, in addition to three-fold rotational symmetry, has three two-fold ($C_2$) rotational symmetry axes about the $a$-axis and every 60° in-plane about the $c$-axis. Using the crystal and experimental symmetries, we determine the indices for the two non-zero independent elements of $\chi^{ED}_{ijk}$: $yyx = yxy = xyy = -xxx$; $yxz = yzx = -xyz = -xzy$

### D. ED SHG under 3 point group

The functional form of the ED SHG intensity under the point group 3 is given by equations (9) and (10) in section III. C. The point group 3 has three-fold rotational symmetry about the out-of-plane $c$-axis ($C_3$). Using the crystal and experimental symmetries, we determine the indices for the six non-zero independent elements of $\chi_{ijk}^{ED}$: $yyy = -xyx = -yxx = -xxy$; $yyz = yzy = xxz = xzx$; $xxx = -xyy = -yxy = -yyx$; $yzx = yxz = -xzy = -xyz$; $zyy = zxx$; $zzz$.

## APPENDIX B: TAUC PLOT MODELING

To demonstrate how additional interband features affect the Tauc plot modeling at low temperatures, we show in Fig. A1 (a) the direct transition Tauc plot for all three materials at 5 K. The nonlinear components of the Tauc plot tend to veer significantly away from the linear regimes. This allows us to separate the Tauc plot into various regions to apply an Urbach tail correction. The Urbach tail corresponds to the exponential decay seen in the absorption spectrum below the band edge of a material and can be caused by a variety of phenomena. Commonly, this Urbach tail arises from phonons, impurities, excitons, and/or structural disorders in a material. Its effect on the absorption spectrum systematically lowers the predicted band gap energy when applying the Tauc model.

To perform this correction in our Tauc plots, we fit the linear regime in the energy range above the prominent peak discussed in section IV. B. The intersection between a linear fit to this region and to that of the band edge corresponds to the estimated band gap energy [43]. In all three materials, this Urbach tail correction adds a relatively constant blue shift of about 8 meV to estimated band gap energies. Additional corrections may be needed for the low temperature interband features. However, as they do not demonstrate a linear trend, we cannot apply the same correction technique as with the Urbach tail.

To support our choice of a direct transition model, we show in Fig. A1 (b) a comparison of the two allowed transition Tauc models for $RbFe(SeO_4)_2$. The functional form is given by Equation (11) and discussed in section IV.B. This shows that the direct model is a better fit to our absorbance data as there is a larger range of energies for where the Tauc plot is linear. The Tauc plots for $RbFe(MoO_4)_2$ and $RbFe(SO_4)_2$ similarly show better agreement with the direct model.

## APPENDIX C: SHG SCANNING MICROSCOPY FOR $RbFe(SeO_4)_2$

Using a 1 μm spot size, we mapped RA SHG patterns at various locations across our $RbFe(SeO_4)_2$ single crystal. While there was apparent spatial inhomogeneity in the SHG signal level, we typically found the same rotational offset and RA SHG patterns with equal-sized lobes as shown above for sites 1 and 2 in Fig. A2 (b). If large inherent strain were present, this would break various symmetries and we would expect variances in the RA SHG pattern, not just the SHG signal level. Strain would most likely manifest in our RA SHG measurements as varying rotational offsets, unequal-sized lobes, or a pattern with a different symmetry entirely. Most selected sites on the crystal face did not have any of these variations in the RA SHG pattern, however there were a few rare sites on the order of 1 μm that seem to break the three-fold symmetry such as site 3 in Fig. A2 (b).

The SHG scanning image is also used to determine the angle correction for the RA SHG polar plots. This angle correction is determined by how far the crystal axes are rotated from the table horizonal, which can be found using the SHG map. The RA SHG polar plot is then rotated accordingly. The same procedure is used with white light imaging.

**APPENDIX D: REFRACTIVE INDEX MEASUREMENTS FOR RbFe(SeO$_4$)$_2$**

Fig. A3 shows the modeled refractive index and extinction coefficient for RbFe(SeO$_4$)$_2$ based on ellipsometry measurements. Measurements were performed on a J.A. Woollam M-2000 Ellipsometer and the complex refractive index was modeled using the associated CompleteEASE software package.

Assumptions when modeling the refractive index included a negative ($n_e < n_o$) uniaxial material, a transparent region for energies 0.77 - 2.48 eV (500 – 1600 nm), and the absence of surface roughness and internal layers beyond the surface. Above 3 eV, the data reveals structure not captured by the UV-VIS transmission absorbance measurements. Using the extinction coefficient shown in Fig. A3 instead of absorbance measurements, we found a band gap energy (3.096 eV) that is in the specified error bars in Fig. 2 (d) for RbFe(SeO$_4$)$_2$ at room temperature. Measured reflectivity levels were typically on the order of 0.5 - 2% of transmission levels for RbFe(MoO$_4$)$_2$ and RbFe(SO$_4$)$_2$ and 5% for RbFe(SeO$_4$)$_2$.

For RbFe(SeO$_4$)$_2$, we estimate the magnitude of the susceptibility tensor to be $d_{22} = \frac{1}{2}\chi_{222} = 0.23$ pm V$^{-1}$ for a fundamental wavelength of 800 nm. This is several orders of magnitude larger than what we might expect for EQ SHG [11] and is comparable to common doubling crystals that have a similar refractive index to RbFe(SeO$_4$)$_2$. Two examples are quartz (α-SiO$_2$) and KDP (KH$_2$PO$_4$) crystals, which have nonlinear optical coefficients of $d_{11} = 0.46$ pm V$^{-1}$ and $d_{36} = 0.63$ pm V$^{-1}$ at 1.060 μm, respectively [31]. The susceptibility strength is determined using the SHG intensity from our RA SHG measurements and fittings to the appropriate point group. The signal is corrected using the magnitude of incoming and outgoing electric fields, which are determined by the experimental set-up. A Fresnel correction is also applied using the refractive index for the extraordinary ray of the material [30].

# FIGURES AND FIGURE CAPTIONS

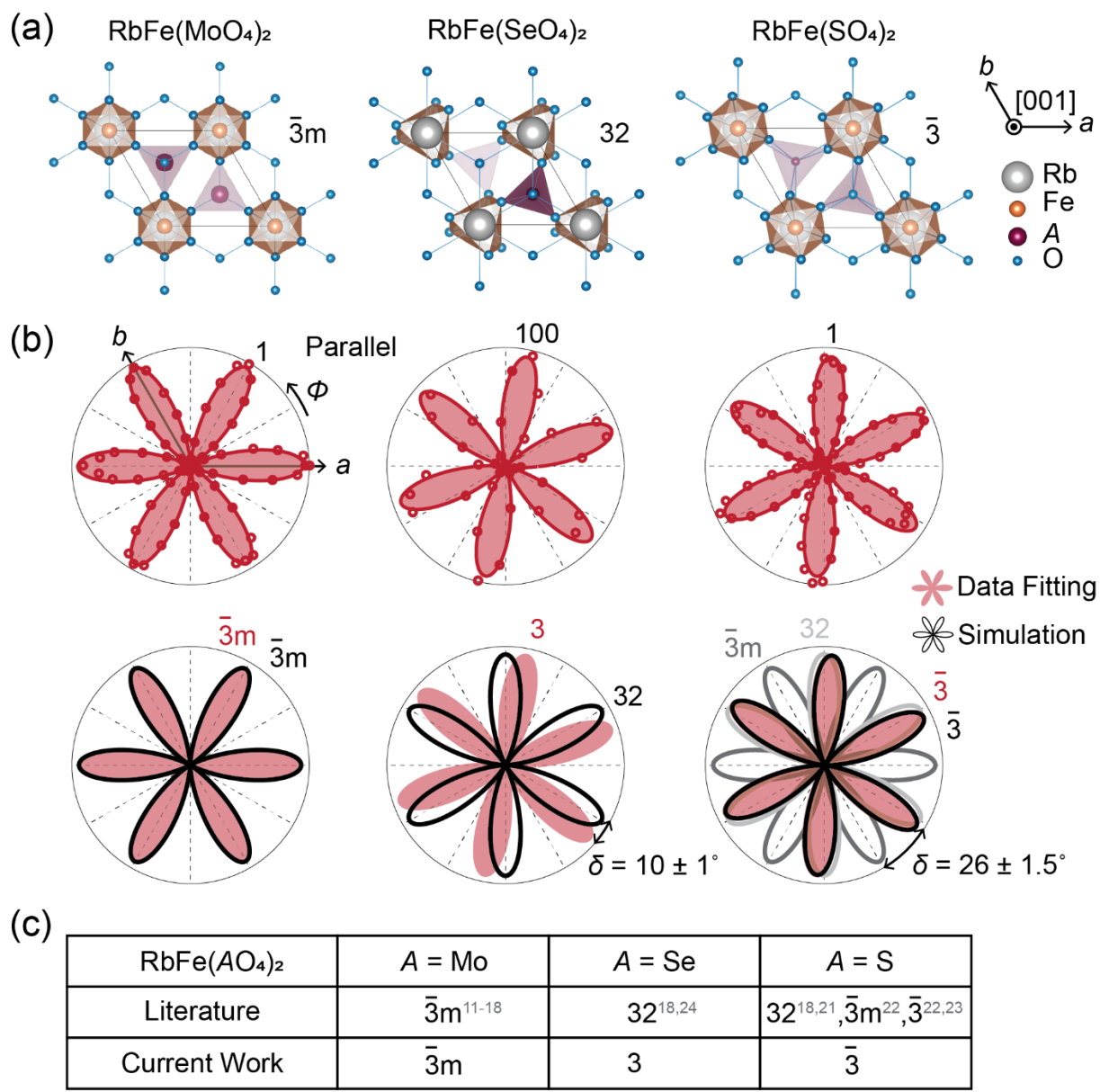

**FIG. 1. (Single column figure, scaled 200%)** (a) Crystal structure of each compound using the literature assigned point group organized by atomic weight of the *A* site with increasing weight towards the right. The in-page plane corresponds to the (001) plane (*ab* plane) and the out-of-page axis corresponds to the *c*-axis of each crystal. (b) RA SHG spectroscopic results of each material. The first row displays measurements from the parallel channel detection scheme. The markers correspond to raw data points and the filled in pattern corresponds to data fitting. The *a*- and *b*-axes locations shown in the first pattern are consistent across each plot. Estimated maximum signal amplitudes are listed next to each pattern and are relative to the signal level of RbFe(MoO$_4$)$_2$. The second row compares fitted data with simulated results for the trigonal point groups proposed in literature. (c) Table comparing the literature assigned point group to the point group used to fit the RA SHG data of each compound.

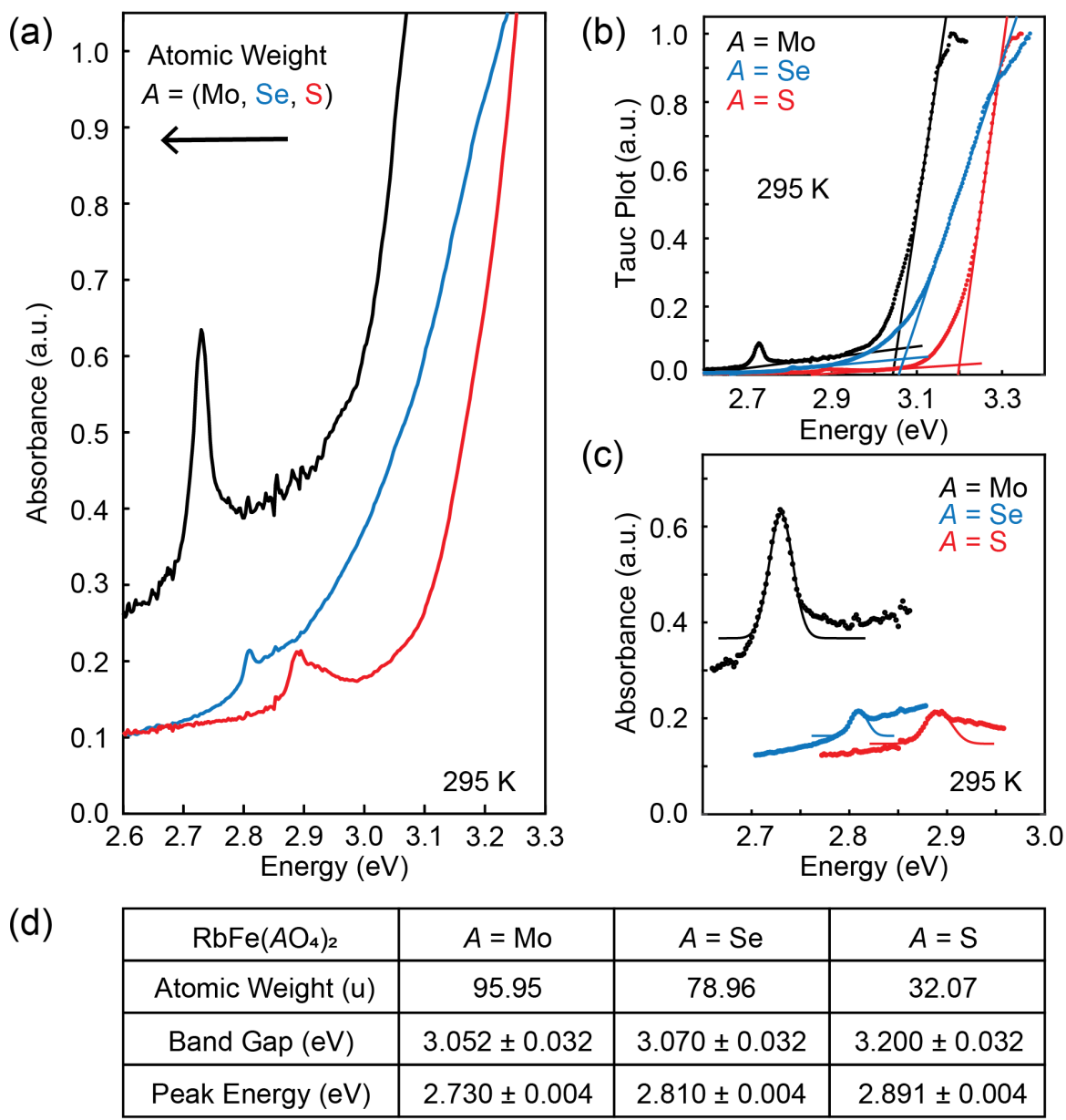

FIG. 2. (Single column figure, scaled 200%) Room temperature UV-VIS measurements and feature fittings for all three compounds. The color scheme is consistent across all panels where black corresponds to RbFe(MoO$_4$)$_2$, blue to RbFe(SeO$_4$)$_2$, and red to RbFe(SO$_4$)$_2$. (a) Absorbance measurements for all three compounds at room temperature. (b) Direct Tauc plot, $(\alpha\hbar\omega)^2 = A(\hbar\omega - E_g)$, versus photon energy, using absorbance measurements as the absorption coefficient to estimate band gap energies for each compound. Markers correspond to absorbance data used in the Tauc equation and the solid lines correspond to the linear fitting (Appendix B). (c) Example of sub-band gap resonance anharmonic oscillator fitting to determine the central energy and FWHM

of the observed features. The contribution of the Urbach tail on the high energy side of each feature is excluded in the fitting. Markers correspond to the absorbance data and the solid lines correspond to gaussian fittings. (d) Estimated room temperature band gap and central feature energies as compared to the atomic weight of the *A*-site. Uncertainty levels for the band gap energy are estimated to be ~ ±1% using protocols from Ref. [35]. Peak energy error bars are determined predominantly from the spectrometer calibration uncertainties, which are correlated to the spectral resolution, as fitting errors are relatively negligible.

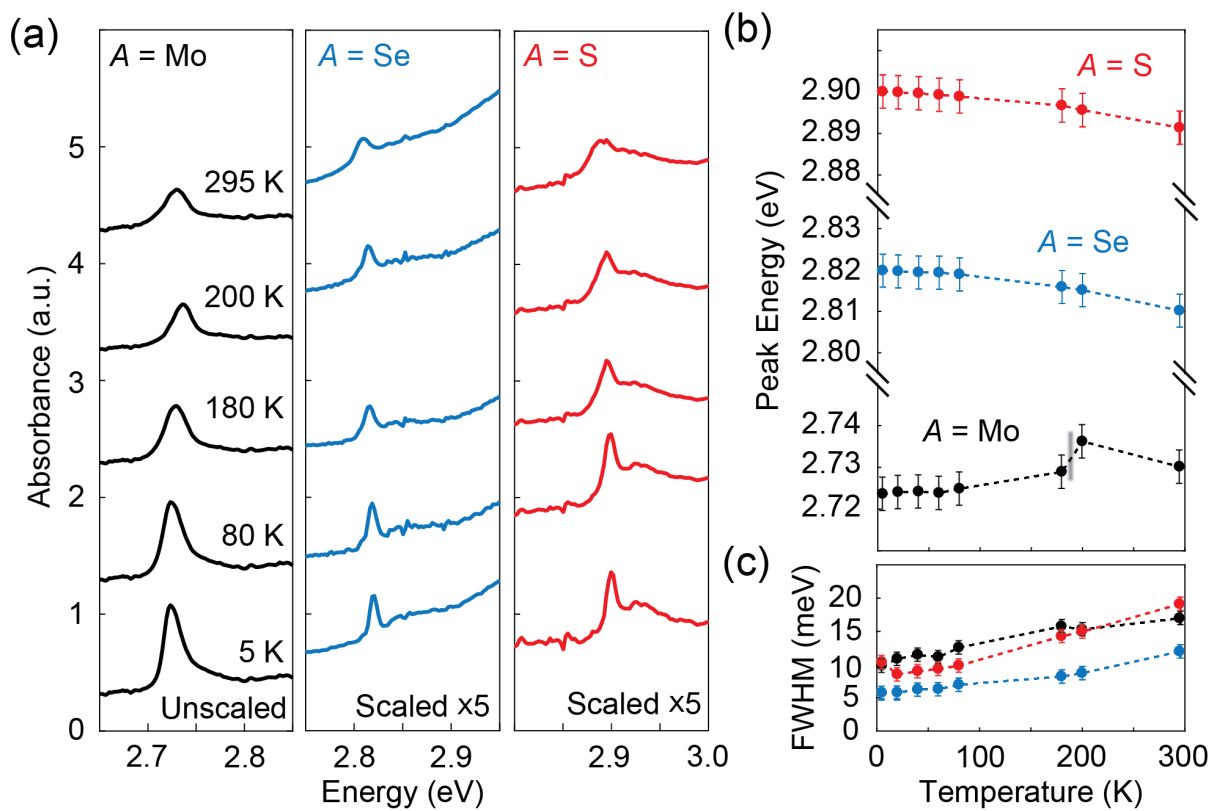

**FIG. 3. (Single column figure, scaled 200%)** (a) Absorbance temperature dependence of the most prominent resonance below the band gap for $RbFe(MoO_4)_2$, $RbFe(SeO_4)_2$, and $RbFe(SO_4)_2$ respectively. Black, blue, and red correspond with compounds $RbFe(MoO_4)_2$, $RbFe(SeO_4)_2$, and $RbFe(SO_4)_2$ respectively in all three panels. (b) Fitted peak energy temperature dependence. The grey bar in the trendline for $RbFe(MoO_4)_2$ indicates the temperature at which the phase transition from $\overline{3}m$ to $\overline{3}$ occurs. Uncertainty levels for the peak energy are again determined from the spectrometer resolution. (c) Fitted resonance full width half maximum (FWHM) temperature dependence. Markers indicated fitted data and dashed lines are present for guidance. For the FWHM, uncertainty levels are determined by the fitting error which is captured in the marker size.

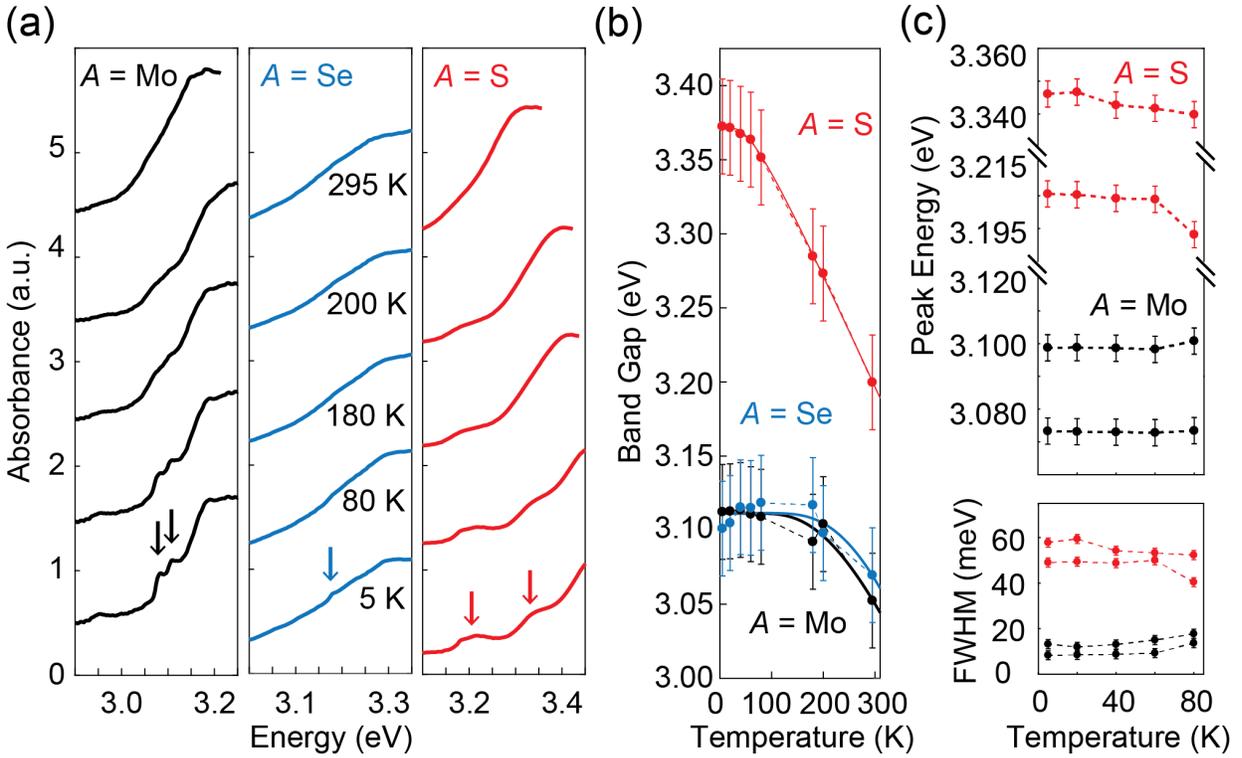

**FIG. 4. (Single column figure, scaled 200%)** (a) Absorbance temperature dependence of the band gap and band gap features. Black, blue, and red correspond to compounds RbFe(MoO$_4$)$_2$, RbFe(SeO$_4$)$_2$, and RbFe(SO$_4$)$_2$ respectively in all three panels. (b) Markers correspond to the band gap temperature dependence from 0 to 295 K determined using the direct Tauc model fitting on the linear regime of the band edge (above any band edge resonances). A correction for the Urbach tail is also employed (Appendix B). The solid line represents a fitting of the data using the thermodynamic model from Equation (12). Again, uncertainty levels for the band gap energy are estimated to be $\sim \pm 1\%$ [35]. (c) Fitted peak energy temperature dependence and FWHM from 0 to 80 K for the peaks indicated by the arrows in panel (a). Only values of resonances present in RbFe(MoO$_4$)$_2$ and RbFe(SO$_4$)$_2$ are listed as the resonance at 3.18 eV for RbFe(SeO$_4$)$_2$ is only distinguishable from the band edge at 5 K. Markers indicate fitted data and dashed lines are present for guidance. Uncertainty levels are determined from the spectrometer resolution and fitting errors for the peak energy and FWHM, respectively.

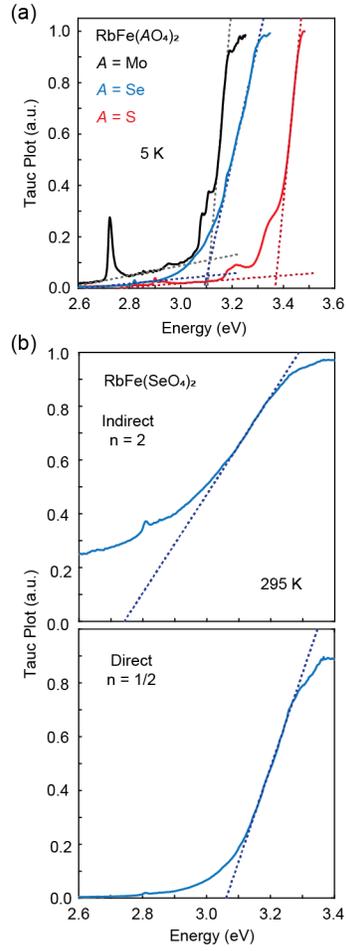

**FIG. A1. (Single column figure, to scale)** (a) Direct transition Tauc plot for RbFe(MoO$_4$)$_2$, RbFe(SeO$_4$)$_2$, and RbFe(SO$_4$)$_2$ at 5 K. (b) Comparison between an indirect and direct Tauc plot for RbFe(SeO$_4$)$_2$ given by the relation $(\alpha\hbar\omega)^{1/n} = A(\hbar\omega - E_g)$ versus photon energy,

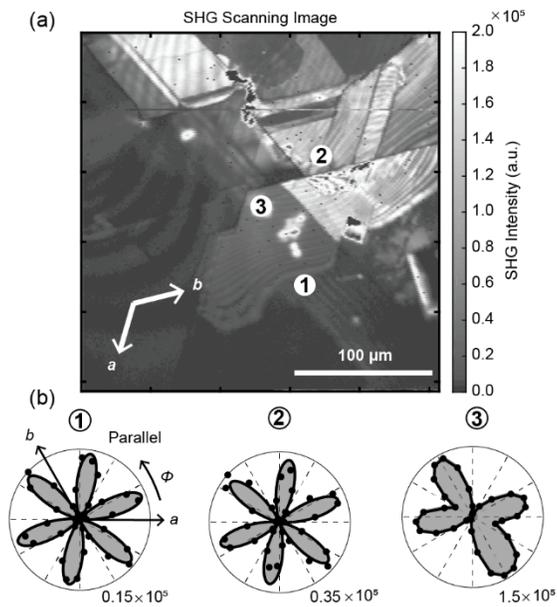

**FIG. A2. (Single column figure, to scale)** (a) SHG scanning image of RbFe(SeO$_4$)$_2$ single crystal. The relative position of the *a*- and *b*- crystal axes are shown in the bottom left corner (b) Polar RA SHG plots at various sites on the crystal. The site locations are given by the corresponding numbers in (a) and (b).

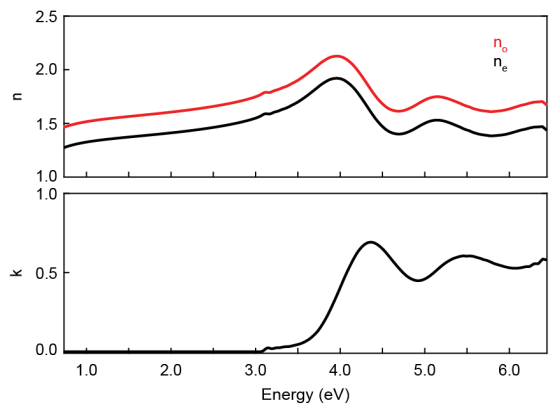

**FIG. A3. (Single column figure, to scale)** Modeled refractive index (*n*) and extinction coefficient (*k*) as a function of energy for RbFe(SeO$_4$)$_2$.